\documentclass[prl,twocolumn,showpacs]{revtex4}
\usepackage{epsfig}
\begin{document}

\author{I.M. Sokolov$^{(1)}$ and J. Klafter$^{(2)}$}
\title{Death of linear response and field-induced dispersion in subdiffusion}
\date{\today}
\affiliation{
$^{(1)}$
Institut f\"{u}r Physik, Humboldt Universit\"{a}t zu Berlin,
Newtonstr. 15, 12489 Berlin, Germany\\
$^{(2)}$ School of Chemistry, Tel Aviv University, Tel Aviv 69978, Israel}
\date{\today}

\begin{abstract}
We discuss the response of continuous time random walks to an oscillating
external field within the generalized master equation approach. We concentrate on the time
dependence of the two first moments of the walker's displacements.
We show that for power law waiting time distributions with $0 < \alpha < 1$
corresponding to a semi-Markovian situation showing nonstationarity the mean particle
position tends to a constant, and the response to the external perturbation dies out.
On the other hand, the oscillating field leads to a new additional contribution to
the dispersion of the particle position, proportional to the square of its amplitude
and growing with time.  These new effects, amenable to experimental observation,
result directly from the non-stationary property of the system.
\end{abstract}

\maketitle

Continuous time random walks (CTRWs) with on-site waiting time distributions being power laws 
lacking the first moment have been shown to provide a powerful tool to describe systems which 
display subdiffusion. These subdiffusive CTRWs are non-Markovian (semi-Markov) 
processes characterized by nonstationarity (aging).
Examples are dispersive charge transport in disordered semiconductors, contaminants 
transport by underground water, motion of proteins through cell membranes and many others 
(see e.g. \cite{MeK,Phys2day,PW} for reviews and popular accounts).
In the absence of time-dependent external perturbations 
CTRW is a process subordinated to simple random walk, thus leading to 
the description within a framework of fractional Fokker-Planck equations 
\cite{MeK,Phys2day,Bar1,sokolov,SKCha}. 
The response to time-dependent fields, which has been less explored, is a much more 
delicate problem, due to nonstationary nature of such CTRW leading to a variety of 
effects related to aging  \cite{Barkai,SBK,SKB,Grigo1,Grigo2}. 
Here we investigate the response of a particle which performs a non-stationary CTRW 
to an external oscillating field by calculation the first two moments of its displacement. 
New effects are observed which are absent in the Markovian case. This should be also 
amenable to experimental observation. 

Let us first describe the model adopted throughout this work. 
In a biased decoupled CTRW a particle arriving to a site $i$ at time $t^{\prime}$
stays there for a sojourn time $t$, distributed with the probability density
function $\psi(t)$ independent of external perturbation. This waiting time PDF is 
considered to follow a power-law $\psi(t) \propto \tau_0^\alpha t^{-1-\alpha}$ where 
$\tau_0$ gives the typical timescale for a jump.
Leaving a site it makes a random step of length $a$ in either direction. 
This step is assumed to be instantaneous on the timescale of typical waiting times and
changes of external parameters; the direction of this step can be biased by the 
time-dependent external force $f(t)$. In what follows we turn to dimensionless units
and measure length in units of $a$ and time in units of $\tau_0$. 
The probabilities of going to the right $w_{i-1,i}(t)$ (from site $i-1$ to site $i$) and to the
left (from site $i+1$ to site $i$) are assumed to be 
\begin{equation}
w_{i-1,i}(t) = \frac{1}{2} + \frac{\mu}{2} f(t); \quad w_{i+1,i}(t) = 
\frac{1}{2} - \frac{\mu}{2} f(t).
\label{bias} 
\end{equation}

Our description of response of CTRW to external fields is based 
on generalized master equation (GME) approach. Several different ways to derive
the corresponding equations are known in the literature 
(e.g. \cite{KlaS,Burshtein,ChvoR,Goy,chechkin}), the one especially fitted 
to describing response to time-dependent fields is given in \cite{Sokolov0}.
For the sake of completeness, we give here the sketch of this derivation. The 
GME follows from two balance conditions,
the probability conservation in a given state and under transitions between different states.

The probability balance for the site $k$ reads 
\begin{equation}
\dot{p}_k=j_k^{+}-j_{k}^{-}(t),
\label{Eq1}
\end{equation}
(where the dot denotes the time derivative) with $j_{k}^{\pm}(t)$ denoting the gain and loss 
currents for a site. A particle leaving its site $k$ at 
time $t$ either was in $k$ from the very beginning or arrived at $k$ at some 
$0<t^{\prime }<t$ so that
\begin{eqnarray}
j_{k}^{-}(t) &=& \psi (t)p_{k}(0)+\int_{0}^{t}\psi (t-t^{\prime})j_{k}^{+}(t^{\prime })dt^{\prime } \\
&=&  \psi (t)p_{k}(0)+\int_{0}^{t}\psi (t-t^{\prime })\left[ \dot{p}_{k}(t^{\prime })+
j_{k}^{-}(t^{\prime })\right] dt^{\prime }, \nonumber 
\end{eqnarray}
where in the second line Eq.(\ref{Eq1}) was used. The formal solution to this equation 
can be expressed through an integro-differential operator
\begin{equation}
j_{k}^{-}(t)= \hat{\Phi} p_k(t) = 
\frac{d}{dt}\int_{0}^{t} M(t-t')p_k(t')dt'
\end{equation}
with the memory kernel $M(t)$ given by its Laplace transform 
\begin{equation}
\tilde{M}(u)= \frac{\tilde{\psi}(u)}{1-\tilde{\psi}(u)}.
\end{equation} 

The probability conservation for transitions between different sites give the relation between 
the gain current in the state $k$ and loss currents at neighboring sites: 
\begin{equation}
j_k^+= w_{k-1,k}(t)j_{k-1}^-+w_{k+1,k}(t)j_{k+1}^-.
\end{equation} 
Inserting the corresponding expressions into the first balance equation gives a GME for $P_k(t)$:
\begin{equation}
\dot{p}_k(t) =  w_{k-1,k}(t) \hat{\Phi} p_k(t) + w_{k+1,k}(t) \hat{\Phi} p_k(t) 
-  \hat{\Phi} p_k(t).
\label{GME}
\end{equation}
Note that the integro-differential operator $\hat{\Phi}$ does not commute with the function 
of time $w_{ij}(t)$.

Using now Eq.(\ref{bias}) for the transition probabilities 
and passing to the continuum limit we get a generalized Fokker-Planck equation
\begin{equation}
\frac{\partial}{\partial t} p(x,t) = \left[ -\mu f(t) \nabla
 + \frac{1}{2} \Delta \right]  \frac{d}{dt} \int_{0}^{t} M(t-t^{\prime }) p(t') dt'.
\label{fineq}
\end{equation}
For Markovian random walk process with exponential waiting time 
distribution (corresponding to $\alpha =1$ and thus to $M(t)= 1$)
this equation reduces to a usual time-dependent Fokker-Planck equation. 
For power-law waiting-time distributions $\psi(t) \propto t^{-1-\alpha}$ with
$0< \alpha <1$ one gets $M(t) \propto t^{\alpha-1}$ and the integro-differential operators on the 
right-hand side of this equation get proportional to the operator of fractional 
Riemann-Liouville derivative $\frac{d}{dt} \int_{0}^{t} M(t-t^{\prime }) f(t') dt' \propto  
\; _0 D_t^{1-\alpha} f(t)$. This is exactly the case we now concentrate on.

The CTRWs with $0<\alpha <1$ are known to show a variety of phenomena connected
with non-stationarity (related also to the so-called aging property 
\cite{Barkai,SBK, SKB,FeigWin,MoBou,Montus,Barkai2}). One of the
manifestations of aging is the decay of response of the system to an alternating or pulsed field 
in course of the time (i.e. when its age grows), see Ref. \cite{SBK}.
Here we consider the response of the system to a sinusoidal external field $f(t)$.
We start from Eq.(\ref{fineq})
and consider the moments of $p(x,t)$, $m_{n}(t)=\int_{-\infty
}^{\infty }x^{n} p(x,t)dx$ of the probability distribution of particle's positions. 
These moments can be easily obtained by multiplying
both sides of Eq.(\ref{fineq}) by $x^{n}$ and integration. Assuming the system to be infinite
and spatially homogeneous we get by partial integration
\begin{equation}
\int_{-\infty }^{\infty }x^{n}\nabla p(x,t^{\prime })dx =  -nm_{n-1}(t^{\prime }) 
\end{equation}
and
\begin{equation}
\int_{-\infty }^{\infty }x^{n}\Delta p(x,t^{\prime
})dx=n(n-1)m_{n-2}(t^{\prime }) 
\end{equation}
(for $n\geq 2$). 
Thus, general equations for the moments are given by 
\begin{equation}
\dot{m}_{n}(t)=n \mu f(t) \hat{\Phi} m_{n-1}(t) +\frac{n(n-1)}{2} \hat{\Phi} m_{n-2}(t). 
\end{equation}
To be able to use these equations in the whole range of $0 \leq m < \infty$ 
one can formally put $m_{0}=1$ and $m_{-1}(t)=0$. 
The equations for the first moment (mean displacement) and the second
(dispersion) read:
\begin{equation}
\dot{m}_{1}(t)= \mu f(t) \hat{\Phi} 1 = \mu f(t) M(t)
\end{equation}
and 
\begin{equation}
\dot{m}_{2}(t) =  2\mu f(t)\hat{\Phi} m_{1}(t)+ M(t).
\end{equation}

Note that for semi-Markovian cases with $0<\alpha<1$ $M(t) \propto t^{\alpha -1}$ is the decaying function
of time. Therefore the response $\dot{m}_{1}(t)$ to the perturbation vanishes in course of
the time leading to the effect we call ``death of linear response'' in
systems showing subdiffusion.

The second moment, $m_{2}(t)=\int_{0}^{t}\dot{m}_{2}(t^{\prime })dt^{\prime}$ 
consists essentially of two contributions, $m_2(t)= \sigma_1^2(t) + \sigma_2^2(t)$, 
the one depending on the external perturbation (through the first moment $m_1$) 
\begin{equation}
\sigma_1^2(t)=2 \int_0^t dt' \mu f(t')\frac{d}{dt'} \int_{0}^{t'} M (t'-t'')m_{1}(t'')dt''
\end{equation}
and of the field-independent purely (sub)diffusive contribution 
\begin{equation}
\sigma_2^2(t) =  \int_0^t M(t') dt'. 
\label{sigma2}
\end{equation}

Let us discuss the overall
structure of expressions for the first and the second moment for the case of a periodic force
$f(t)=f_0 \sin \omega t$. The first moment is:
\begin{eqnarray}
m_1(t) & = &\mu f_0 \int_0^t dt' \sin \omega t' M(t') dt' \nonumber \\
& = & \int_0^t dt' \frac{e^{i \omega t'}
-e^{-i \omega t'}}{2i} M(t') dt'.
\end{eqnarray}
Turning to the Laplace domain we get
\begin{equation}
\tilde{m}_1(u) = \frac{\mu f_0 }{2iu}\left[ \tilde{M}(u-i\omega) - \tilde{M}(u+i\omega)\right]
\end{equation}
as it follows from the shift theorem. The asymptotic behavior of the first moment
then follows straightforwardly. For $u \rightarrow 0$ (corresponding to $t \rightarrow \infty$)
we have
\begin{equation}
\tilde{m}_1(u) = \frac{\mu f_0 }{2iu}\left[\tilde{ M}(-i\omega) -\tilde{ M}(i\omega)\right]
\end{equation}
being the Laplace transform of a constant $m_1(\infty) = -\mu f_0 \mathrm{Im} \tilde{M}(-i \omega)$.
The Laplace transform of the field-dependent contribution to the 
second moment (again obtained by using the shift theorem) reads:
\begin{eqnarray}
\tilde{\sigma}_1^2(u) =&& -\frac{\mu^2 f_0^2}{4u} 
\left\{\tilde{M}(u-i\omega)\left[\tilde{M}(u-2i\omega)-\tilde{M}(u) \right]- \right. \nonumber \\
&& \left. \tilde{M}(u + i\omega)\left[\tilde{M}(u)-\tilde{M}(u+2 i \omega) \right]\right\}.
\end{eqnarray}
To obtain the asymptotic behavior of $\tilde{\sigma}_1^2(u)$ it is enough to note that for power-law
waiting time distributions with $0 < \alpha <1$, $\tilde{M}(u) \propto u^{-\alpha}$ which diverges when
$u \rightarrow 0$. Thus, the leading contribution to $\tilde{\sigma}_1^2(u)$ is
\begin{eqnarray}
\tilde{\sigma}_1^2(u) & \simeq & \mu^2 f_0^2 \frac{\tilde{ M}(-i\omega) \tilde{M}(u) +
\tilde{M}(i\omega)\tilde{ M}(u)}{4u} \nonumber \\
& = & \mu^2 f_0^2 \mathrm{Re} \tilde{M}(i \omega) 
\frac{\tilde{M}(u)}{u}.
\label{sigma12}
\end{eqnarray}
This means that the asymptotic growth of the field-dependent contribution to the dispersion 
is given by $\sigma_1^2(t) \propto \mu^2 f_0^2 \int_0^t M(t') dt' \propto  \mu^2 f_0^2 t^\alpha $.

The most important feature of this result is that although the first moment of the
distribution stagnates, the field-dependent contribution to the dispersion  
continues growing, a manifestation of a new effect, specific for non-stationary CTRWs, 
namely the field-induced dispersion. This growing contribution is absent for the Markovian and 
asymptotically Markovian processes  ($\alpha = 1$) only due to the fact that the 
corresponding prefactor $\mathrm{Re} (i \omega)^{-\alpha}$ vanishes. 

To get an impression on the overall behavior of the first and the second moments let us 
consider the special case $\alpha = 1/2$. In our numerical example we set $\omega =1$. In this case
\begin{eqnarray}
\dot{m}_{1}(t)& = & \mu f_{0}\sin t\frac{d}{dt}\int_{0}^{t}(t-t^{\prime
})^{-1/2}dt^{\prime }\\
& = & \Gamma(1/2) \mu f_0 \sin t \;_0 D_t^{1/2}1 =
\mu f_{0}t^{-1/2}\sin t.  \nonumber 
\end{eqnarray} 
Note that the fractional derivative of the constant does not vanish. 
From this expression we see that the susceptibility of the system to an external
force decays in course of the time as $t^{-1/2}$; namely its linear response dies out.
Integrating this expression over time we get the analytical form for $m_{1}(t)$
\begin{equation} 
m_{1}(t)=\mu f_0 \sqrt{2\pi }S(\sqrt{t})
\end{equation}
involving the Fresnel sinus-integral $S(x)=\left( 1/\sqrt{2\pi }\right)
\int_{0}^{x}dt\sin \left( t\right) /\sqrt{t}$.
The Fresnel integral tends to a constant value for large values of its 
argument; the final value of the first moment is mostly determined by the
value of the external perturbation at short times, when the system was ``young'',
the result of what was called ``Freudistic'' memory of aging systems
in Ref.\cite{SKB}. The behavior of $m_1(t)$ is shown in Fig. 1. 
\begin{figure}
\centering
\scalebox{0.5}{\includegraphics{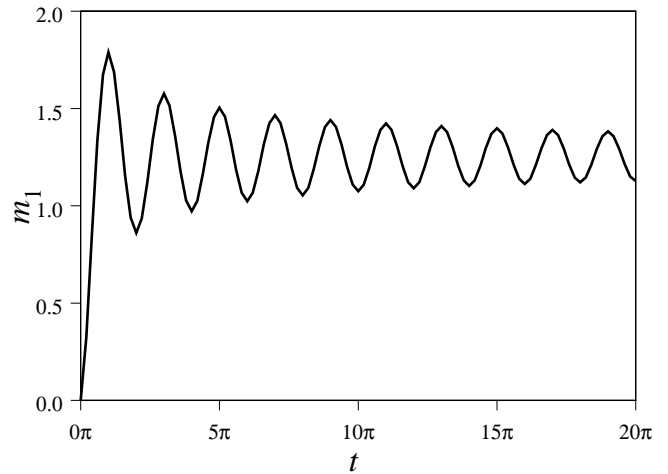}}
\caption{The mean displacement $m_1(t)$ (measured in units of $\mu f_0$) in a 
CTRW model with $\alpha=1/2$ as a function of time for $\omega =1$. Note that 
for $t$ large the displacement stagnates: The linear response of the system 
to external sinusoidal field dies.}
\end{figure}

The behavior of the field-dependent contribution to the second moment is
\begin{equation}
\sigma_1^2(t) = \mu f_{0}\int_{0}^{t}dt^{\prime }\sin t^{\prime }\frac{d}
{dt^{\prime }}\int_{0}^{t^{\prime }}(t^{\prime }-t^{\prime \prime
})^{-1/2}m_{1}(t^{\prime \prime })dt^{\prime \prime }.
\end{equation}
Integrating this expression by parts twice we get:
\begin{equation}
\sigma_1^2(t) = 2 \mu^2 f_{0}^{2}\left[ \sin (t)\Psi (t)-\int_{0}^{t}\Psi (t^{\prime
})\cos (t^{\prime })dt^{\prime }\right] .
\label{sigma21}
\end{equation}
with 
\begin{equation}
\Psi (t)=\int_{0}^{t}\sqrt{(t-t^{\prime })/t^{\prime }}\sin t^{\prime
}dt^{\prime }
\end{equation}
The numerical evaluation of the integral is shown in Fig.2 showing
$[\sigma_1^2(t)]^{1/\alpha}$, i.e. the square of the corresponding expression, Eq.(\ref{sigma21}).
The leading contribution $t^\alpha$ to the overall behavior corresponds to the overall linear
growth of $[\sigma_1^2(t)]^{1/\alpha}$.
Interesting is the subleading asymmetric oscillatory behavior overplayed
on this overall growth. The amplitude of these oscillations decays on a linear
plot, but seems to stay constant in the one presenting the square of the function.
This means that the decay of the subleading term follows essentially $t^{-1/2}$.
The field-independent subdiffusion contribution $\sigma_2^2(t)$ grows as $t^{1/2}$ 
according to Eq.(\ref{sigma2}).  

\begin{figure}
\centering
\scalebox{0.5}{\includegraphics{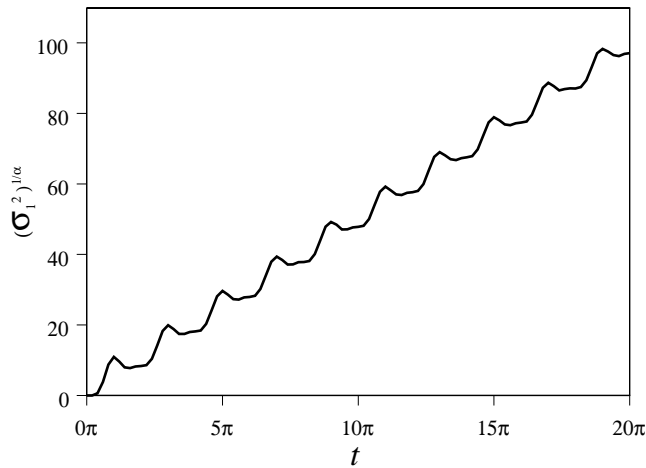}}
\caption{
Shown is the square of  $\sigma_1^2(t)$ (measured in units of $\mu^2 f_0^2$)
as a function of $t$  for $\alpha=1/2$ and $\omega=1$.} 
\end{figure}

Let us summarize our findings.
We discussed the behavior of a particle performing continuous-time random walks with a 
power-law distribution of waiting times lacking the first moment ($\psi(t) \propto t^{-1-\alpha}$ with 
$0< \alpha <1$) under the influence of oscillating external field. Using the approach based on the
generalized master equation we derive equations for the first two moments of the displacement.
The first moment of the displacement stagnates, an effect we term ``death of linear response''. 
The second moment, on the contrary, grows as $t^\alpha$ and contains, in addition to the normal 
(sub-)diffusion contribution, a field-induced contribution, proportional to the square of the 
external field. This new effect which shows up in non-stationary CTRW is absent 
in the Markovian case ($\alpha =1$) since the corresponding prefactor vanishes.

\end{document}